# Optical Devices for Cold Atoms and Bose-Einstein Condensates

Naceur Gaaloul[1,2], Amine Jaouadi[1,2], Mourad Telmini[2], Laurence Pruvost[3], and Eric Charron[1]

[1] *Laboratoire de Photophysique Moléculaire du CNRS,*
*Bâtiment 210, Université Paris-Sud, 91405 Orsay cedex, France*

[2] *Laboratoire de Spectroscopie Atomique, Moléculaire et Applications,*
*Faculté des Sciences de Tunis, 2092 Tunis, Tunisia*

[3] *Laboratoire Aimé Cotton du CNRS,*
*Bâtiment 505, Université Paris-Sud 11, 91405 Orsay Cedex, France*

**Abstract.** The manipulation of cold atoms with optical fields is a very promising technique for a variety of applications ranging from laser cooling and trapping to coherent atom transport and matter wave interferometry. Optical fields have also been proposed as interesting tools for quantum information processing with cold atoms. In this paper, we present a theoretical study of the dynamics of a cold $^{87}$Rb atomic cloud falling in the gravity field in the presence of two crossing dipole guides. The cloud is either deflected or split between the two branches of this guide. We explore the possibilities of optimization of this device and present preliminary results obtained in the case of zero-temperature dilute Bose-Einstein condensates.

**Keywords:** cold atoms, optical manipulation, beam splitter, deflector.
**PACS:** 03.75.Be, 32.80.Pj, 32.80.-t, 39.25.+k.

## INTRODUCTION

The manipulation of cold atoms with optical fields is a well developed technique in atom optics [1]. Far off resonant optical fields can be used not only to trap cold atoms, but also to guide them or even to implement more sophisticated devices such as atomic beam splitters for instance [2,3]. The implementation of such devices is motivated by the implementation of interferometry experiments with cold atoms. In particular, recent experiments [4] have shown that very high contrast interference patterns can be obtained even with thermal sources of atoms. In this article, we present a theoretical study of the dynamics of cold atoms falling in the gravity field in a configuration similar to the one of Ref. [2]. In this experimental study, a thermal cloud of cold atoms at $T \simeq 10-20\,\mu$K is released in the gravity field in the presence of two red-detuned far off-resonant laser beams. The first beam is oriented vertically, and forms an angle $\gamma$ with the second one (see Fig.1 for a schematic representation).

These two beams create a guiding potential for the atoms given by the expression

$$V_{\text{guide}}(\vec{r}) = \frac{\hbar\Gamma}{2} \frac{1}{4|\delta|/\Gamma} \frac{I(\vec{r})}{I_s}, \quad (1)$$

where $\Gamma$, $\delta$ and $I_s$ denote the natural linewidth, the detuning and the saturation intensity of the $5S_{1/2} \rightarrow 5P_{3/2}$ $^{87}$Rb atomic transition. The atomic coordinate $\vec{r} \equiv (x, z)$ is restricted to the plane formed by the two crossing dipole guides in this study, and $I(\vec{r})$ is the spatial distribution of light intensity. In the following, we consider two ideal Gaussian laser beams of intensities $I_0$ and $I_1$, and waists $w_0$ and $w_1$, with

$$I(\vec{r}) = I_0 e^{-\frac{2x^2}{w_0^2}} + I_1 e^{-\frac{2x'^2}{w_1^2}}, \qquad (2)$$

where the rotated coordinate $x'$ is shown Fig.1. The atoms are therefore subjected to a global non-separable guiding potential of the following form

$$V_{guide}(x, z) = -U_0 e^{-\frac{2x^2}{w_0^2}} - U_1 e^{-\frac{2[x\cos\gamma + (z-z_0)\sin\gamma]^2}{w_1^2}}, \qquad (3)$$

where the potential depths $U_0$ and $U_1$ are simply proportional to $I_0$ and $I_1$.

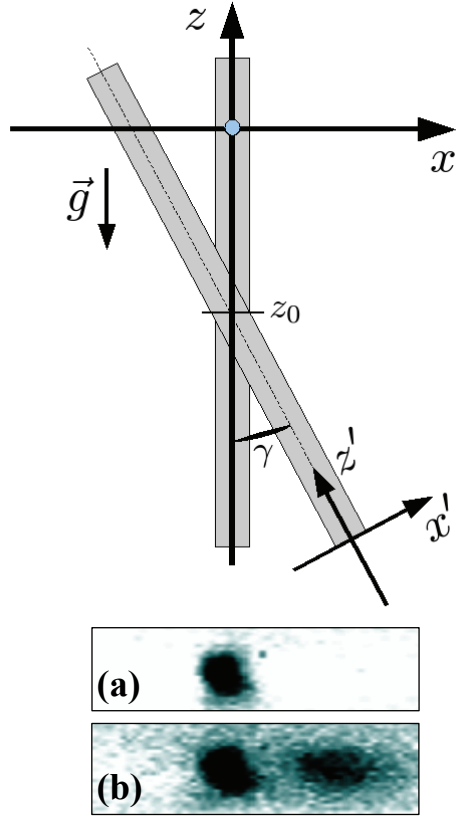

**FIGURE 1.** Schematic representation of the optical device and fluorescence images of the atoms at the height $z_P = -1$ cm $\leq z_0$: (a) Vertical guide only, (b) Complete beam splitter setup. See Ref. [2].

When the oblique beam is switched off ($U_0 \neq 0$ and $U_1 = 0$), the atoms are confined and guided in the vertical guide only (Fig.1(a)). For $U_0 \neq 0$ and $U_1 \neq 0$ a splitting of the atomic cloud is observed (Fig.1(b), see Ref. [2] for details).

# THEORETICAL FRAMEWORK AND NUMERICAL RESULTS

## Theoretical Model

To study this splitting dynamics, we solve numerically the time dependent Schrödinger equation for the atomic motion

$$i\hbar \frac{\partial}{\partial t}\Psi(x,t) = -\frac{\hbar^2}{2m}\frac{\partial^2}{\partial x^2}\Psi(x,t) + V_{eff}(x,t)\Psi(x,t), \quad (4)$$

where $m$ is the $^{87}$Rb atomic mass and $\Psi(x,t)$ denotes the time-dependent translational atomic wave function.

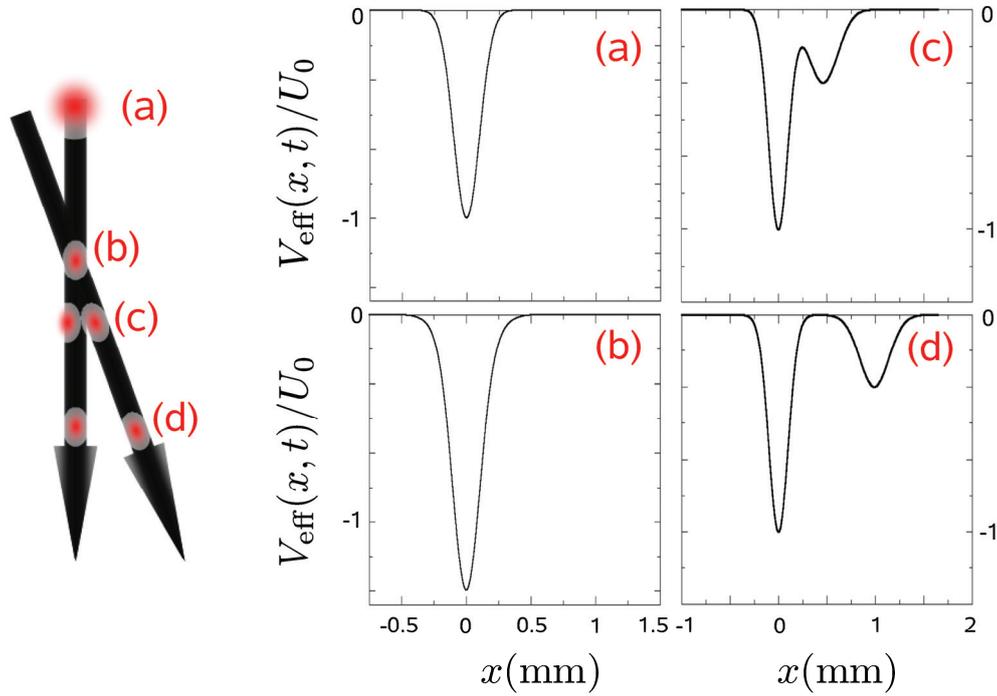

**FIGURE 2.** Effective one-dimensional potential $V_{eff}(x,t)$ at various heights $z = -\frac{1}{2}gt^2$. The graphs (a)-(d) correspond to the labels shown on the left picture. The potential parameters are $U_0 = 30$ μK, $U_1 = 10$ μK, $w_0 = 0,3$ mm, $w_1 = 0.45$ mm, $z_0 = -4$ mm, and $\gamma = 10$ deg.

The presence of the gravity field, and therefore of the two-dimensional atomic motion, is taken into account by the introduction of an effective time-dependent potential given by

$$V_{eff}(x,t) = V_{guide}(x, z = -\tfrac{1}{2}gt^2), \quad (5)$$

which assumes a simple free-fall dynamics in the $z$-direction. This effective potential is shown at different times $t$ (0, 29, 40, and 45 ms) in Fig. 2. The validity of this simple one-dimensional approach has already been verified numerically [5].

We assume that the atom is initially ($t = 0$, $z = 0$) in a well defined vibrational level $v = v_0$ of the guiding potential

$$V_0(x) = -U_0\, e^{-\frac{2x^2}{w_0^2}} \qquad (6)$$

created by the vertical laser beam. We then propagate the translational wave packet $\Psi(x,t)$ using the splitting technique of the short-time propagator [5,6]. At the end of the propagation, the wave function $\Psi(x,t_f)$ is analyzed to determine the efficiency of the beam splitter.

## Beam Splitter and Deflector for Cold Atoms

A typical coherent splitting over both potential wells can be seen in Fig. 3 for a given initial state ($v_0 = 6000$ in the present case). This figure shows the initial and final atomic probability distributions $|\Psi(x,0)|^2$ and $|\Psi(x,t_f)|^2$ as well as the associated effective one-dimensional potentials.

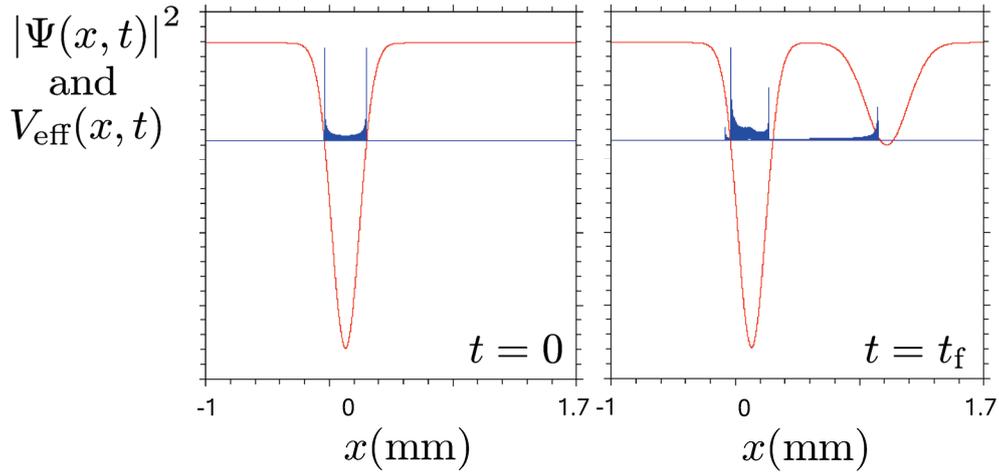

**FIGURE 3.** Atomic probability distributions at times $t = 0$ and $t = t_f = 45$ ms for $v_0 = 6000$. The potential parameters are the same as in Fig. 2.

The probability of finding the atom in the oblique dipole guide at the end of the propagation can be evaluated for each initial trap state $v_0$. The total splitting efficiency of the present beam splitter setup is then obtained by averaging this state-dependent probability over the initial distribution of vibrational levels calculated using a Maxwell-Boltzmann assumption [5]. The variation of this splitting efficiency is shown Fig. 4 as a function of the potential depth ratio $U_1/U_0$.

One can notice here that the splitting efficiency increases monotonically with $U_1$. For large values of $U_1$, an almost complete deflection of the atomic cloud can be realized. A completely symmetric splitting is also predicted when $U_1 \simeq 1.1 \times U_0$. This last prediction is in agreement with the experiment [7].

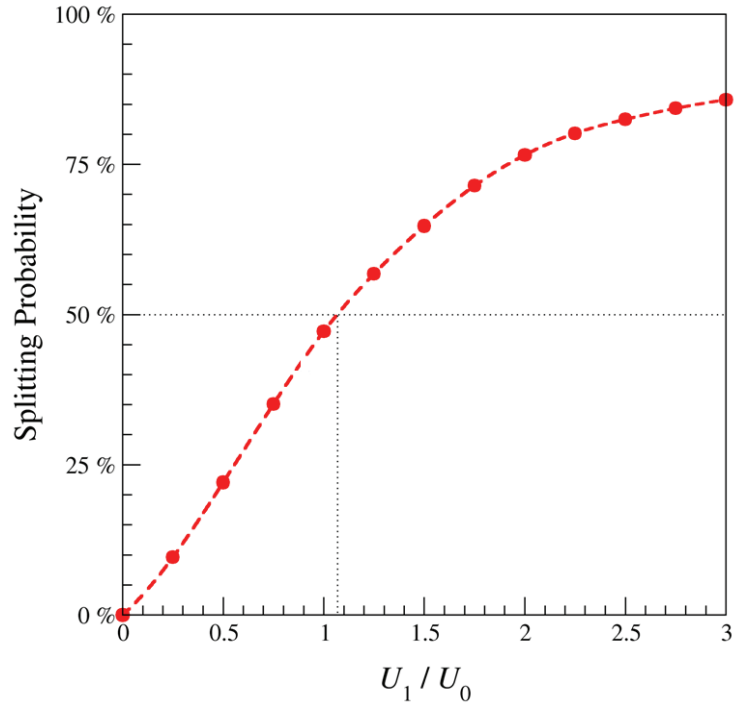

**FIGURE 4.** Splitting efficiency as a function of $U_1/U_0$. The parameters are the same as in Fig. 2. The temperature of the initial atomic cloud is $T = 14\ \mu K$.

The results shown in this figure indicate that a high degree of control exists in this type of experimental configurations since the splitting efficiency can be modified almost at will. The possibility of controlling the splitting efficiency is one of the very important keys for future cold atom interferometry experiments, especially because the fringe contrast in such experiments is strongly dependent on the numerical value of the transmission and reflection coefficients of the beam splitter.

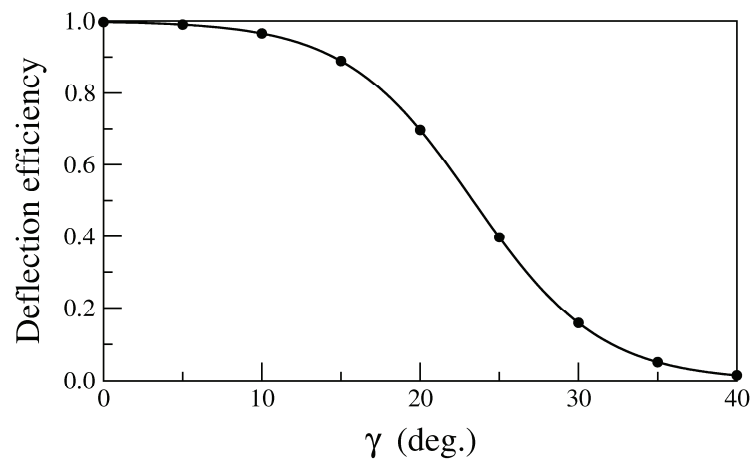

**FIGURE 5.** Deflection efficiency as a function of $\gamma$. The potential parameters are the same as in Fig. 2, except for $U_1 = 4\,U_0 = 120\ \mu K$. The temperature of the initial atomic cloud is $T = 10\ \mu K$.

In addition, if the vertical guide is switched off at time $t_0 = \sqrt{2|z_0|/g}$, when the atoms reach the crossing point ($z = z_0$), a deflector device is easily implemented. Figure 5 shows effectively that for reasonable values of laser intensities and detunings, corresponding to $U_1 = 4\,U_0 = 120\,\mu K$, an efficient deflection is obtained over a large range of deflection angles $0 \leq \gamma \leq 25$ deg. For large values of the ratio $U_1/U_0$, the deflector efficiency is indeed very close to 100%, and in this case, the deflector setup can be regarded as an atomic mirror, which could constitute one of the necessary devices for a future implementation of a Mach-Zehnder–type atom interferometer.

## Application to Bose-Einstein Condensates

Preliminary calculations show that Bose-Einstein condensates are also easily deflected in a similar setup. To perform these simulations, we first solve the time-independent Gross-Pitaevskii equation

$$-\frac{\hbar^2}{2m}\Delta\Phi(\vec{r}) + V_{\text{guide}}(\vec{r})\,\Phi(\vec{r}) + Ng_{2D}\left|\Phi(\vec{r})\right|^2 \Phi(\vec{r}) = \mu\,\Phi(\vec{r}) \qquad (7)$$

where $N$, $\mu$ and $\Phi(\vec{r})$ are the particle number, the chemical potential and the macroscopic condensate wave function (order parameter), respectively. The averaged two-dimensional mean-field interaction term is proportional to the $^{87}$Rb s-wave scattering length $a_0$, following

$$g_{2D} = 2\hbar^2 \left(2\pi\frac{\omega_y}{m}\right)^{1/2} a_0, \qquad (8)$$

where a strong confinement (trapping frequency $\omega_y$) has been assumed in the perpendicular $y-$direction [8]. Equation (7) is solved using the imaginary time relaxation technique [9].

The evolution of the ground state chemical potential $\mu$ with the boson number $N$ is shown Fig. 6(a), with the simple expectation value obtained from the two-dimensional Thomas-Fermi approximation [10]. This approximation is obtained by neglecting the kinetic energy term in the Gross-Pitaevskii equation (7). It is usually accurate for large particle numbers and repulsive interactions.

If we approximate the Gaussian light potential by the usual harmonic expression

$$V_H(r) = -U_0 + \frac{1}{2}m\omega^2 r^2 \qquad (9)$$

with

$$\omega = \frac{2}{w_0}\sqrt{\frac{U_0}{m}}, \qquad (10)$$

one is effectively left with an analytic expression for the condensate density

$$\left|\Phi_{TF}(\vec{r})\right|^2 = \frac{\mu_{TF} - V_H(\vec{r})}{N\,g_{2D}}, \qquad (11)$$

and the normalization condition on $\Phi_{TF}(\vec{r})$ yields a simple relation between the two-dimensional Thomas-Fermi chemical potential $\mu_{TF}$ and the atom number $N$

$$\mu_{TF} + U_0 = \frac{2}{w_0}\left(\frac{g_{2D}U_0}{\pi}N\right)^{\frac{1}{2}}. \qquad (12)$$

Figure 6(a) shows that $\mu + U_0 \rightarrow \frac{1}{2}\hbar\omega$ when $N \rightarrow 1$ and $\mu + U_0 \propto N^{1/2}$ for large values of $N$, as expected.

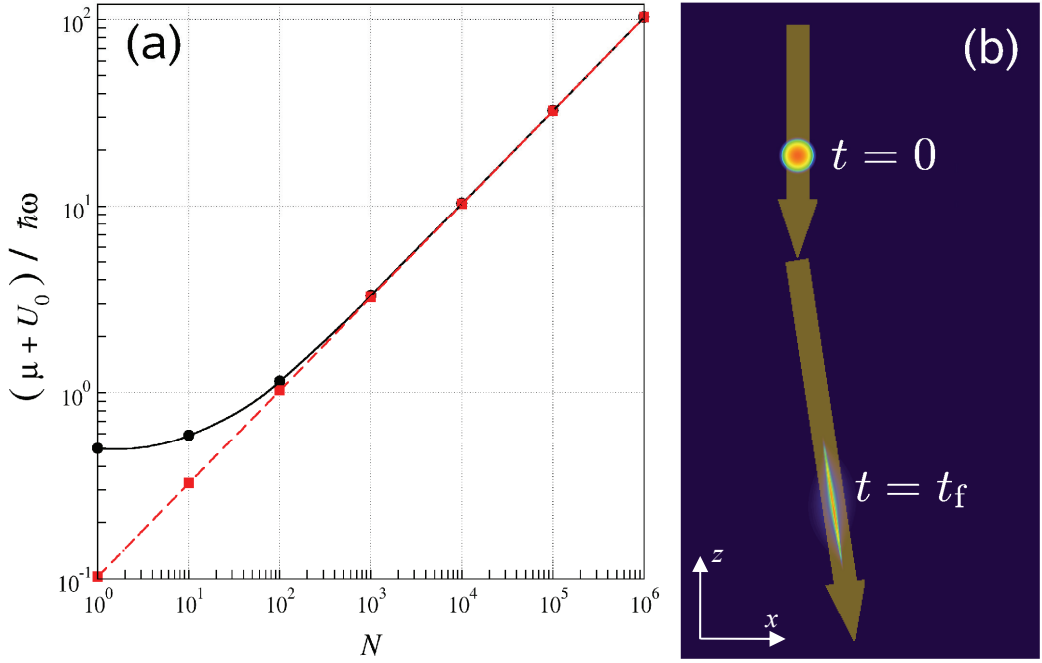

**FIGURE 6.**
(a) Variation of the chemical potential with the boson number. The solid line shows the two-dimensional numerical result while the dashed line represents the expectation value of the Thomas-Fermi approximation. The laser parameters correspond to $U_0 = 10\,\mu K$ and $w_0 = 30\,\mu m$. The strong perpendicular confinement is fixed such that $\omega_y = 10\,\omega$.

(b) Condensate density $|\Phi(\vec{r},t)|^2$ at times $t = 0$ (upper isotropic shape) and $t = t_f \simeq 600\,\mu s$ (lower elongated shape). The potential parameters are $U_0 = 2.4\,\mu K$, $U_1 = 8.0\,\mu K$, $w_0 = 11.3\,\mu m$, $w_1 = 5.0\,\mu m$, $z_0 = -1.0\,\mu m$, and $\gamma = 10\,\deg$.

The condensate dynamics is then followed by solving the time-dependent Gross-Pitaevskii equation

$$-\frac{\hbar^2}{2m}\Delta\Phi(\vec{r},t) + \left[V_{guide}(\vec{r}) + mgz\right]\Phi(\vec{r},t) + Ng_{2D}|\Phi(\vec{r},t)|^2\Phi(\vec{r}) = i\hbar\frac{\partial}{\partial t}\Phi(\vec{r}), \qquad (13)$$

using the same efficient propagation technique [6].

When submitted to a gravity field deflector setup similar to the one of Fig. 1, the initial Bose-Einstein condensate (represented in Fig. 6(b) near the label $t = 0$) starts to fall and expand in the $z-$direction, before being deflected. It is then nicely guided by the oblique laser beam. This effect is shown in Fig. 6(b), which presents the condensate density at time $t = t_f = 600\,\mu s$ in the oblique guide.

## CONCLUSION

To conclude, we have presented a theoretical study of the dynamics of thermal and coherent ensemble of cold atoms in a simple optical beam splitter or deflector device. The effect of gravity has been taken into account, and we have shown that a significant control of the deflection or splitting efficiency is achievable simply by tuning the laser intensity.

## ACKNOWLEDGMENTS

Laboratoire de Photophysique Moléculaire and Laboratoire de Spectroscopie Atomique, Moléculaire et Applications are associated through the CNRS LOTAMP contract. The IDRIS supercomputer center provided computational time under Project No. 08/051848. This work was partially supported by the LRC of the CEA, under Contract No. DSM-0533. Laboratoire de Photophysique Moléculaire and Laboratoire Aimé Cotton are associated with Université Paris-Sud. During the realization of this work, Naceur Gaaloul was financially supported by a grant from the Tunisian Ministry of Higher Education, Research and Technology, and Amine Jaouadi also acknowledges a financial grant from the Université Paris-Sud.